\documentclass[preprint]{aastex}

\shorttitle{A New Cataclysmic Variable in Hercules}
\shortauthors{Price et al.}

\begin{document}
\title{A New Cataclysmic Variable in Hercules}

\author{A. Price, B. Gary, J. Bedient, L. Cook, M. Templeton, C. Pullen, D. Starkey, T. Crawford, R. Corlan, S. Dvorak, 
K. Graham, R. Huziak, R. James, D. Messier, N. Quinn, D. Boyd, J. Blackwell, G. Walker, M. Mattei, D. Rodriguez, M. Simonsen}

\affil{American Association of Variable Star Observers, Clinton B. Ford Astronomical Data and Research Center, 25
Birch Street, Cambridge, MA 02138
\email{\small aavso@aavso.org}
}

\author{A. Henden}
\affil{Universities Space Research Association/U. S. Naval Observatory, Flagstaff, AZ 86001 USA
\email{\small aah@nofs.navy.mil}
}

\author{T. Vanmunster}
\affil{Center for Backyard Astrophysics - Belgium Walhostraat 1A, B-3401 Landen, Belgium
\email{\small Tonny.Vanmunster@cbabelgium.com}
}

\author{P. Garnavich}
\affil{Department of Physics, 225 Nieuwland Science Hall, University of Notre Dame, Notre Dame, IN 46556
\email{\small pgarnavi@miranda.phys.nd.edu}
}

\author{J. Pittichov\'{a}}
\affil{University Hawaii, Institute for Astronomy, 2680 Woodlawn Drive, Honolulu, HI 96822
\email{\small jana@ifa.hawaii.edu}
}

\author{T. Matheson, P. Challis, R. P. Kirshner, E. Adams}
\affil{Harvard-Smithsonian Center for Astrophysics, 60
Garden Street, Cambridge, MA 02138
\email{\small tmatheson@cfa.harvard.edu, pchallis@cfa.harvard.edu, rkirshner@cfa.harvard.edu, eadams@cfa.harvard.edu}
 }

\author{T. Harrison}
\affil{Department of Astronomy, New Mexico State University, 1320 Frenger Mall, Las Cruces, NM 88003
\email{\small tharriso@nmsu.edu }
}

\author{M. D. Koppelman}
\affil{Department of Physics \& Astronomy, University of Minnesota
Minneapolis, MN 55455
\email{\small michael@aps.umn.edu}
}

\author{G. E. Sarty}
\affil{University of Saskatchewan
Saskatoon, Saskatchewan, S7N 5A5, Canada
\email{\small gordon.sarty@usask.ca}
}

\author{D. E. Mais}
\affil{Dept. Astronomy, Palomar College, San Marcos, CA
\email{\small dmais@ligand.com}
}

\begin{abstract} 
 We present time-series observations, spectra and archival outburst data 
of a newly-discovered variable star in Hercules, Var Her 04. Its orbital period, mass ratio, and 
outburst amplitude resemble those of the UGWZ-type subclass of UGSU dwarf novae. However, its supercycle and 
outburst light curve defy classification as a clear UGWZ. Var Her 04 is most similar to the small group 
of possible hydrogen-burning ``period bouncers'', dwarf novae that have passed beyond the period 
minimum and returned. 
\end{abstract}

\objectname[Var Her 04]{EQ J183926+260409}
\object[AAVSO 1835+25]{AAVSO 1835+25}
\objectname{Var Her 04}

\section{Introduction}

 Var Her 04 (EQ J183926+260409; $\alpha$=18:39:26.154 $\delta$=+26:04:09.89 $\pm$ 18mas [J2000]; GCVS designation not assigned yet) is a newly discovered cataclysmic variable in Hercules.  It was discovered at
photographic (Tri-X) magnitude 11.5 by Yuji Nakamura, Kameyama, Mie-ken, Japan on June 13.632 UT \citep{gre04}. Announcement of the discovery was made on June 22 by the AAVSO after
consultation with the IAUC.\footnote{The AAVSO offers two announcement services for these types of objects. The AAVSO customizable MyNewsFlash Service operates automatically
continuously. The AAVSO Alert Notice is a more formal announcement of confirmed events. Both can be accessed and subscribed to at \url{http://www.aavso.org/}} The first known
observation was found in the ASAS-3 survey database \citep{poj02} on June 15.2462 when it was measured at {\it V}= 12.03 $\pm$ 0.03. The first American Association of Variable Star
Observers (AAVSO)-coordinated observation was made on June 23.677.

  Var Her 04 is the NW component of a double star separated by 0.99 $\pm$ 0.04'' (Fig. 1).   Astrometry was measured using the 1.55-m USNO Flagstaff Station (NOFS) telescope. 
The SE component matches the position of 2MASS
J18392619+2604087 ($\alpha$=18:30:26.195 $\delta$=+26:04:09.07 $\pm$ 40mas [J2000]) within its margin of error. The SE component has {\it B-V}=2.09  and
likely contaminates all observations made without filters or in {\it R$_c$} and {\it I$_c$} by a system that cannot separate the two. This includes all observations in
this paper unless otherwise noted. Var Her 04's location
falls within the error margins of the location of 1RXS J183927.1+260409, a bright X-Ray source in the ROSAT catalog (average hardness ratio 0.59 +/- 0.26) \citep{vog99}.

\section{Observations}

 AAVSO observers made 7,551 CCD measurements of the object over 16 days following the announcement of the discovery.  Each observer reduced their own data using comparison stars
published on an AAVSO chart. They are also instructed to synchronize their imaging computer clocks with the USNO NTP time server. Initial photometric calibration of the field was 
done by B. Gary 
from Hereford, Arizona over multiple nights with a 0.36m telescope, ST-8XE CCD and
{\it BVR$_c$} filters at prime focus (Table 1). Comparison stars were chosen based on brightness and {\it B-V} to avoid variability associated with M class stars. Later, 
precision
photometry was obtained over multiple nights using the USNO-FS 1.0m telescope along with a large set of Landolt standards \citep{lan92} having a range of color and airmass. The
field was observed over multiple nights with the University of Hawaii 2.2-meter on Mauna Kea. These images were calibrated with the comparison stars from Henden's photometry. Using
point spread function fitting, Henden also obtained precise astrometry for both the variable and companion along with photometry of the latter.

 The variable was observed at the 1.8m Vatican Advanced Technology
 Telescope on 2004 June 24 (UT). CCD data was taken in the
 Johnson $V$ filter continuously for 5 hours. The majority
 of the images were 10 second exposures (plus a 30 second read-out).
 The data were bias corrected and flat-fielded using twilight
 sky images.

 The optical light curve (Fig. 4) is characterized by a steady decline of 0.17 magnitudes per day until June 24.5000 (JD 2453181.00) when an inflection in the light curve gave rise
to a gentle hump that spanned 0.1 magnitudes over 2 days. On June 26.5 (JD 2453182.00) the light curve resumed a slow decline until June 30.0358 (JD 2453186.55) when a rapid
decline began. The decline resulted in a drop of 1.1 magnitude. The last observation of the decline was made 0.5062 days after it began. The star remained steady around {\it V}=16 
through the end of observations in early
September.  Afterwards, {\it BVR$_c$I$_c$JHK$_p$} photometry was obtained during quiescence providing a measurement of {\it V} = 17.095  $\pm$ 0.029 (Table 2). Precision photometry of the companion revealed 
some brightening 
during the time period that the CV was declining and then it too went into a state of quiescence (Table 3).

 A blue spectrum was obtained on 2004 June 23.32 UT and again on 
2004 June 25.45 with the FAST spectrograph \citep{fab98} at the Cassegrain focus of
the 1.5-meter Tillinghast telescope at F. L. Whipple Observatory on Mt. Hopkins.  The FAST
spectrograph uses a 2688$\times$512 Loral CCD with a spatial scale of
1{\farcs}1 per pixel in the binning mode used for these observations.
The exposure time for the observation was 360s.  The slit was oriented
with a position angle of 70$^{\circ}$; this was not the optimal parallactic angle \citep{fil82}, but
the low airmass (1.01) and wide slit (3$^{\prime\prime}$) imply that
differential slit losses were insignificant.  Standard CCD processing
and optimal spectral extraction were done with IRAF\footnote{IRAF is
distributed by the National Optical Astronomy Observatories, which are
operated by the Association of Universities for Research in Astronomy,
Inc., under cooperative agreement with the National Science
Foundation.}.  Custom routines were used to calibrate the flux data
using the sdO comparison star BD $+28^{\circ}4211$ \citep{sto77} in the
region 3400-4500 \AA\ and the sdG comparison star BD $+26^{\circ}2606$
(Oke \& Gunn 1983) in the region 4500-7500 \AA.  Telluric absorption
features were removed using the intrinsically featureless spectrum of
BD $+26^{\circ}2606$ \citep{wad88,mat00}.

The spectrum (Fig. 2) shows a blue continuum with
double-peaked Balmer lines.  The full width at half maximum of the
H$\alpha$ line is 2000 km s$^{-1}$ and it shows some change in shape between the
 two nights. The Balmer lines become dominated by absorption
 at the blue end of the spectrum. Weak, double-peaked emission
 lines of HeI are detected as well as a NaI absoprtion feature.
 Very broad bumps identified as HeII and high ionization states
 of carbon and nitrogen, generally associated with Wolf-Rayet
 features are also seen. The spectrum is similar to that
 of WZ~Sge during outburst \citep{bro80}.

\section{Superhumps}

 Superhumps were observed from the beginning of AAVSO observations on June 23.3677 (2453179.8677) to June 30.5592 (2453187.0592) (Fig. 5). Before combining data for statistical
analysis, each observer's data set was individually transformed to a uniform zero-point by subtracting a linear fit from each night's observations. This was done so that we could
remove the overall trend of the outburst, and combine all observations into a single data set. We used a date-compensated discrete Fourier transform \citep{fer81} combined with the
CLEANest algorithm \citep{fos95} for period analysis. Superhumps appeared in the earliest time series data set beginning on June 23.3677. Analysis of the observations until June
30.5592 (2453187.0592)  give a superhump period of 0.05778 $\pm$ 0.000001 days (Fig. 6). Then a period of 0.056855 $\pm$ 0.000069 days emerged in quiescence (Fig. 7). The power
spectrum in quiescence is also more complicated. This is probably because of noise from the faint photometry and influence by the red companion.  Further time series CCD
observations spread out over the next ten days revealed no significant change in the power spectrum.

\section{Archival Data}

 The blue RH series patrol plates at the Harvard College Observatory Plate Stacks were inspected for previous outbursts on plates 
dating from 1929 to 1950. 
Four outbursts were detected and one more possible outburst was detected at the plate limit (Table 4). The All Sky Automated Survey 3 (ASAS-3) does not detect the star during
almost nightly observations from March 30.4 2003 to June 6.3 2004. A search through the orphan files in the NSVS database \citep{woz04} revealed no 
detected outbursts from April 5.5 1999 to March 23.5 2000.

\section{Analysis}

 The orbital period, outburst amplitude and 
lack of a recently detected outburst\footnote{An outburst of this object within the last 20 years should have been 
detected by amateurs - as the 2004 outburst was - due to its prime location, brightness and duration 
of outburst. Outbursts during solar conjunction, though, could not
have been observed.} suggest classification as a UGWZ dwarf nova.
However, Var Her 04 displayed 
more frequent and consistent outburst behavior from 1929 - 1950 and there is an absence of the expected UGWZ echo 
outbursts in the 2004 outburst. 

 If $P_{\mbox{sh}} = 0.057780 \pm 0.000001$ days is the superhump period and
$P = 0.056855 \pm 0.000069$ days is the orbital period then we derive a
superhump-period excess, $\epsilon = (P_{\mbox{sh}} - P_{\mbox{orb}})/P_{\mbox{orb}} = 0.016 \pm
0.003$  and an approximate mass ratio of $q = 0.072 \pm 0.0113$ as derived from
\begin{equation}
\epsilon = \frac{0.23 q}{1 + 0.27 q}
\end{equation}
(Patterson 1998).

 There are other objects that display similar erratic outburst behavior with 
orbital periods and mass ratios similar to Var Her 04: EG Cnc, WZ Sge, AL Com and DU UMa \citep{pat98}. This small group 
of stars are believed to be hydrogen-burning ``period bouncers'', dwarf novae that have periods which are lengthening after
evolving through the period minimum. These are theoretically the most common type of CV, but also the most 
elusive as a result of their inherently faint nature and infrequent outbursts. The most obvious common trait of these stars 
is very low $\epsilon$, which is dependant on mass ratio. The low mass ratio implies a very small secondary which 
provides a hint towards the relative age of the system with older systems more likely to have survived a period bounce.

 The faint red companion is close to the variable making it difficult to obtain precision photometry and astrometry of both 
during quiescence. Photometry of this companion suggests a hint of brightening that is coincident with the outburst of the variable 
and beyond the expected photometric error. The online POSS I/II averaged coordinates were checked for proper motion but 
they consist of the averaged position of the red and blue plates. The former will be contaminated by the companion so proper 
motion could not be estimated. Additionally, direct parallax measurement
would be quite difficult because of the companion but may be possible in three years with a large aperture telescope.  
The change in position angle should be detectable in a few years if the stars are part of the same system.
If so, then 
the companion may 
have been reflecting outburst light from the variable.

\section{Conclusions}

      AAVSO observations and study of recent and past Var Her 04 superoutbursts reveal an enigmatic system with an 
outburst light curve and supercycle unlike any other known object. Var Her 04 is most similar to the small group of now five possible 
UGSU hydrogen-burning ``period bouncers''. Searches for past outbursts in other archives and careful monitoring of current 
behavior is needed to establish supercycle behavior evolution. AAVSO observers will carefully monitor the star 
over the next few years for a new outburst. Precision photometry, X-ray observations, and quality spectra 
in quiescence would be useful in firmly establishing the mass ratio of the system. In addition, future proper motion studies of the 
companion using large aperture telescopes is needed to determine any relationship with the faint red companion.

\acknowledgments
      We acknowledge the use of SIMBAD operated through the {\it Centre de Données 
Astronomiques} (Strasbourg). We thank Alison Doane, Curator of the Harvard College Observatory 
Plate Stacks. We also thank John Greaves, Michael Richmond, Dan Taylor, Bob Stine, 
David Cornell, Ron Royer, Dave Hurdis, Jerry Horne, Richard Kinne and Pierre De Ponthierre for various help with 
the AAVSO observing campaign. We acknowledge Mike Calkins for spectrum from the Tillinghast telescope.
       The telescopes used by Huziak and Sarty were made available by Stan Shadick of the 
University of Saskatchewan Physics Department.
       The Digitized Sky Survey (DSS) was produced at the Space Telescope Science Institute under U.S. Government grant NAG W-2166.

\newpage

\figcaption[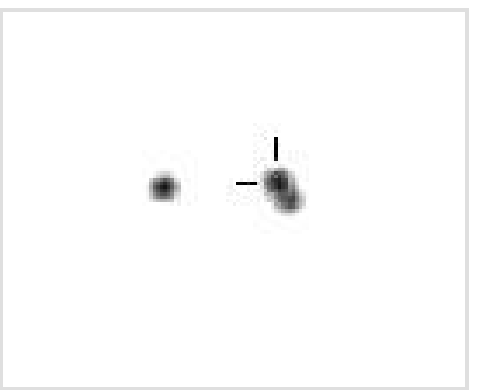]{Var Her 04 (NW component) and companion in R$_c$ at quiescense. Image taken on Aug. 15, 2004 with the University of Hawaii 2.2m Telescope (0.2arcsec/pixel; N up, E right; image fov: 10").
Photometry of Ver Her 04 in quiescence is contaminated by the companion. \label{one}}
\figcaption[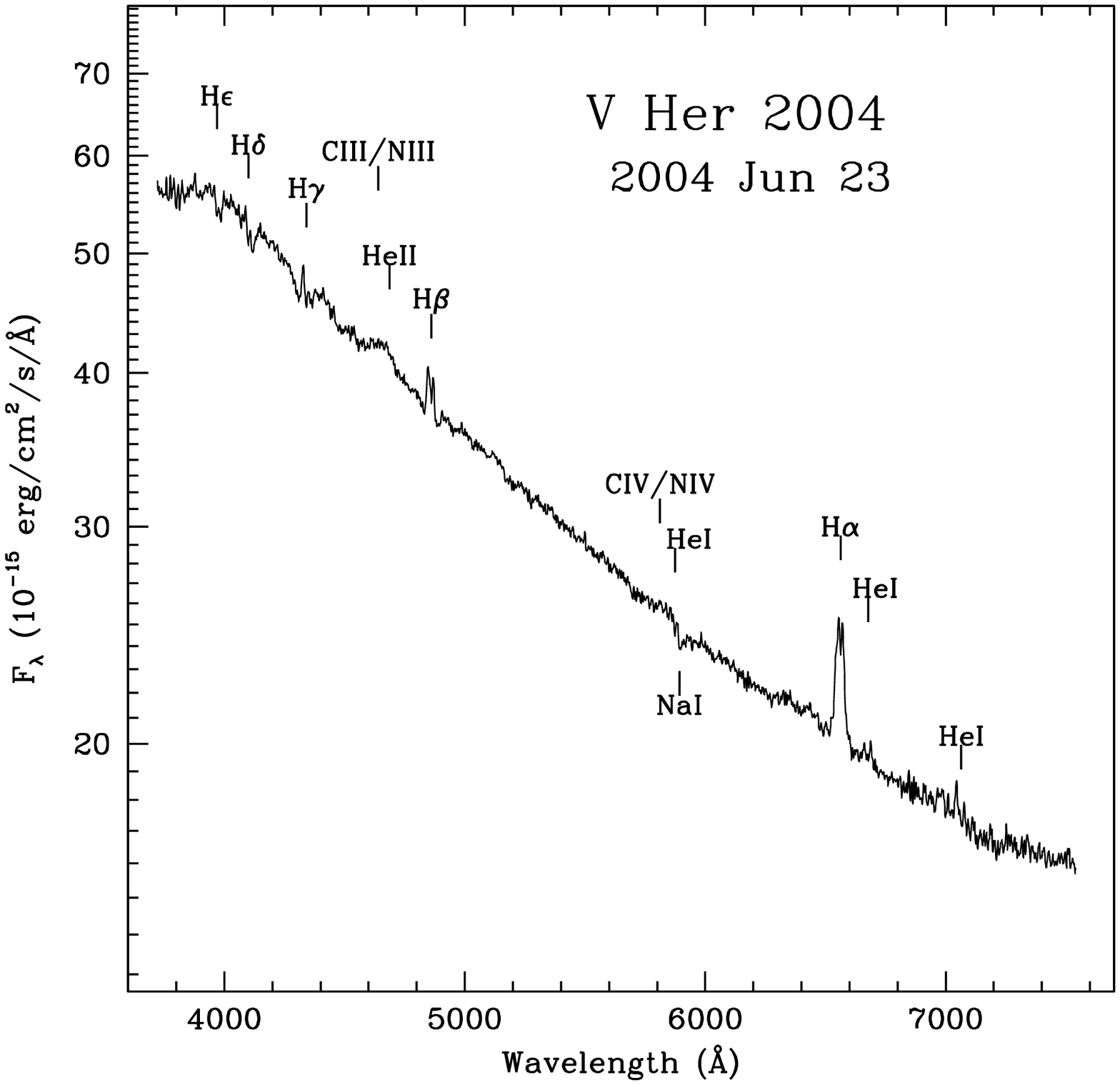]{Blue (370-750nm) spectra from Mt. Hopkins on June 23.32 reveals the typical spectrum of a cataclysmic variable. \label{three}}
\figcaption[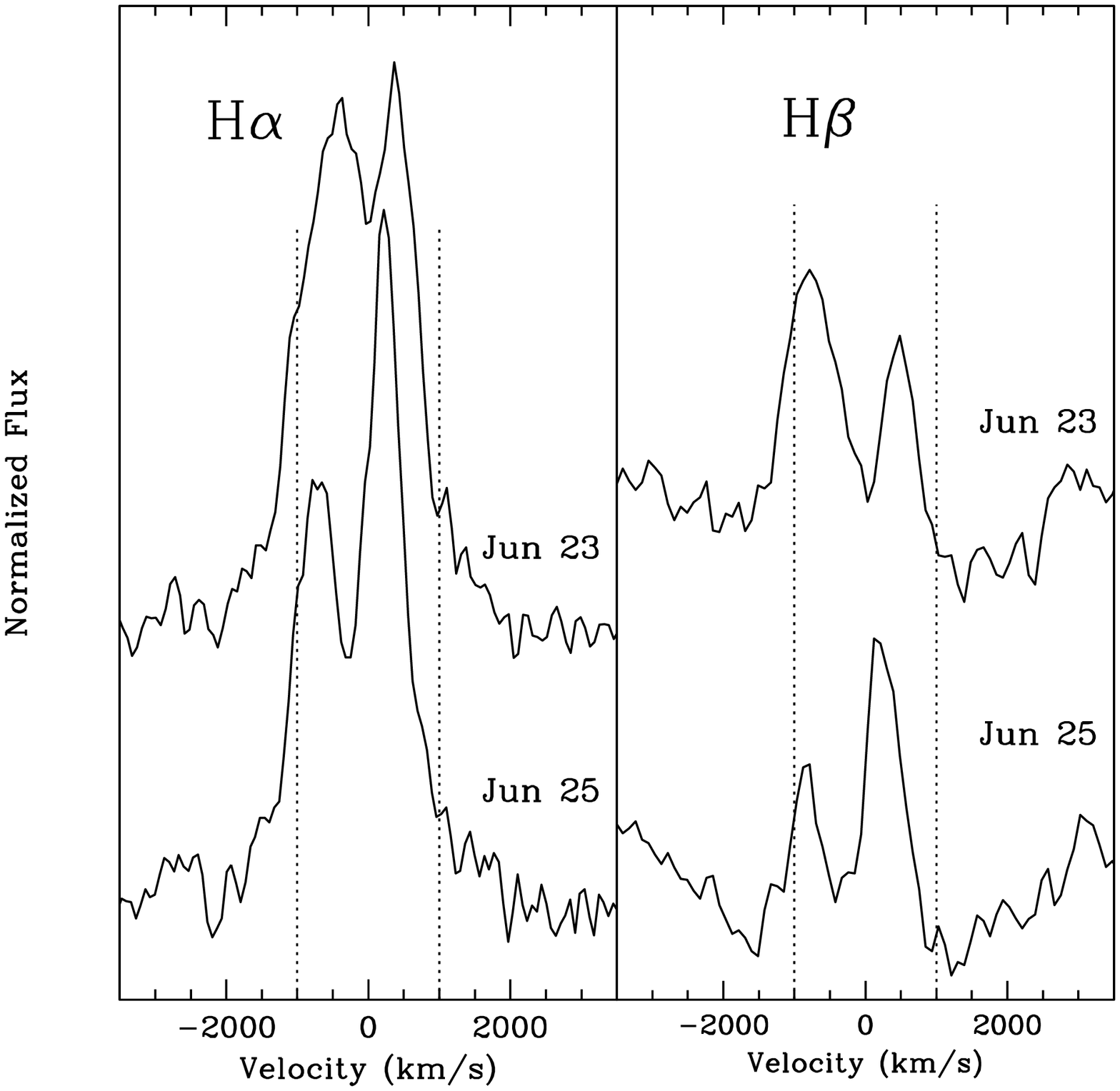]{Comparison of Ha and Hb features in the Mt. Hopkins spectra taken on June 23.32 and June 25.00. \label{four}}
\figcaption[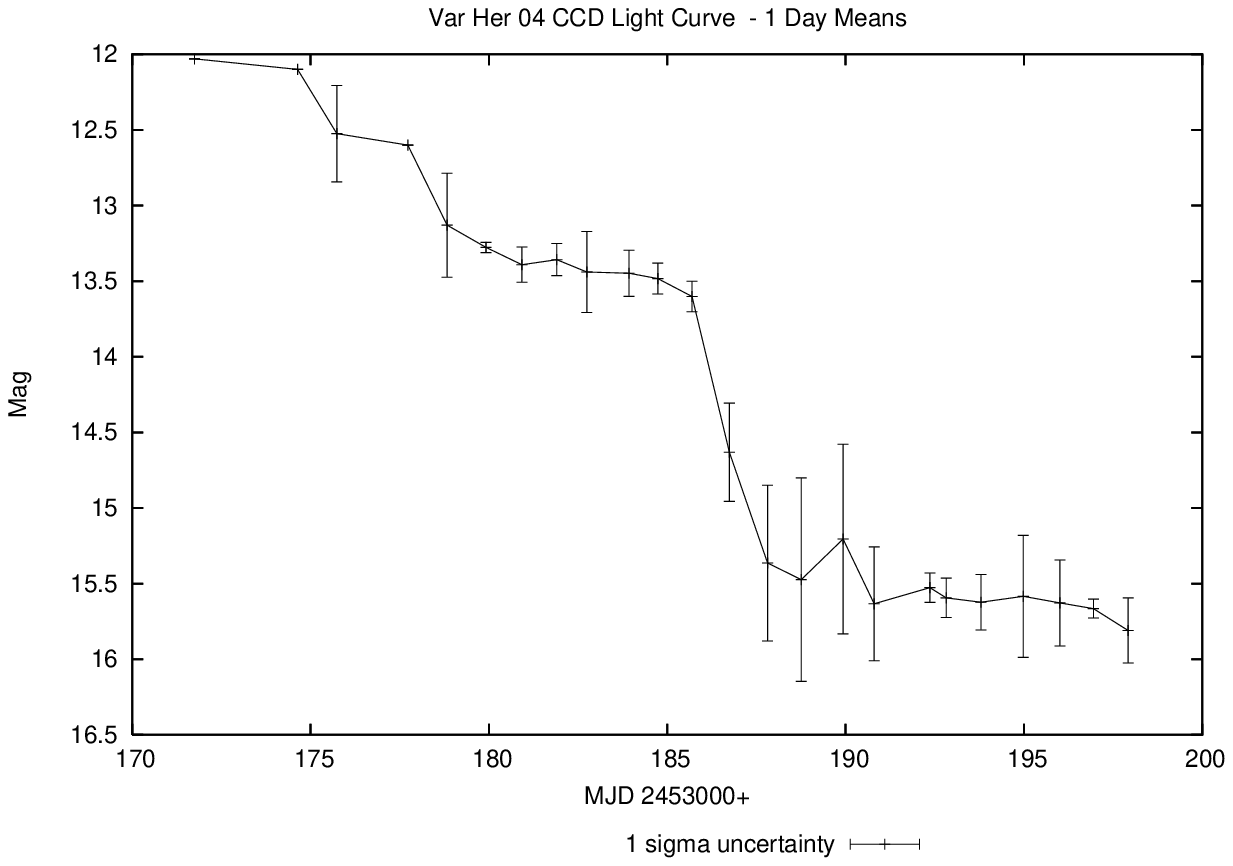]{Var Her 04's Light Curve. Data averaged from AAVSO sources. Combining faint unfiltered observations leads to the larger errors
      at late time. \label{five}}
\figcaption[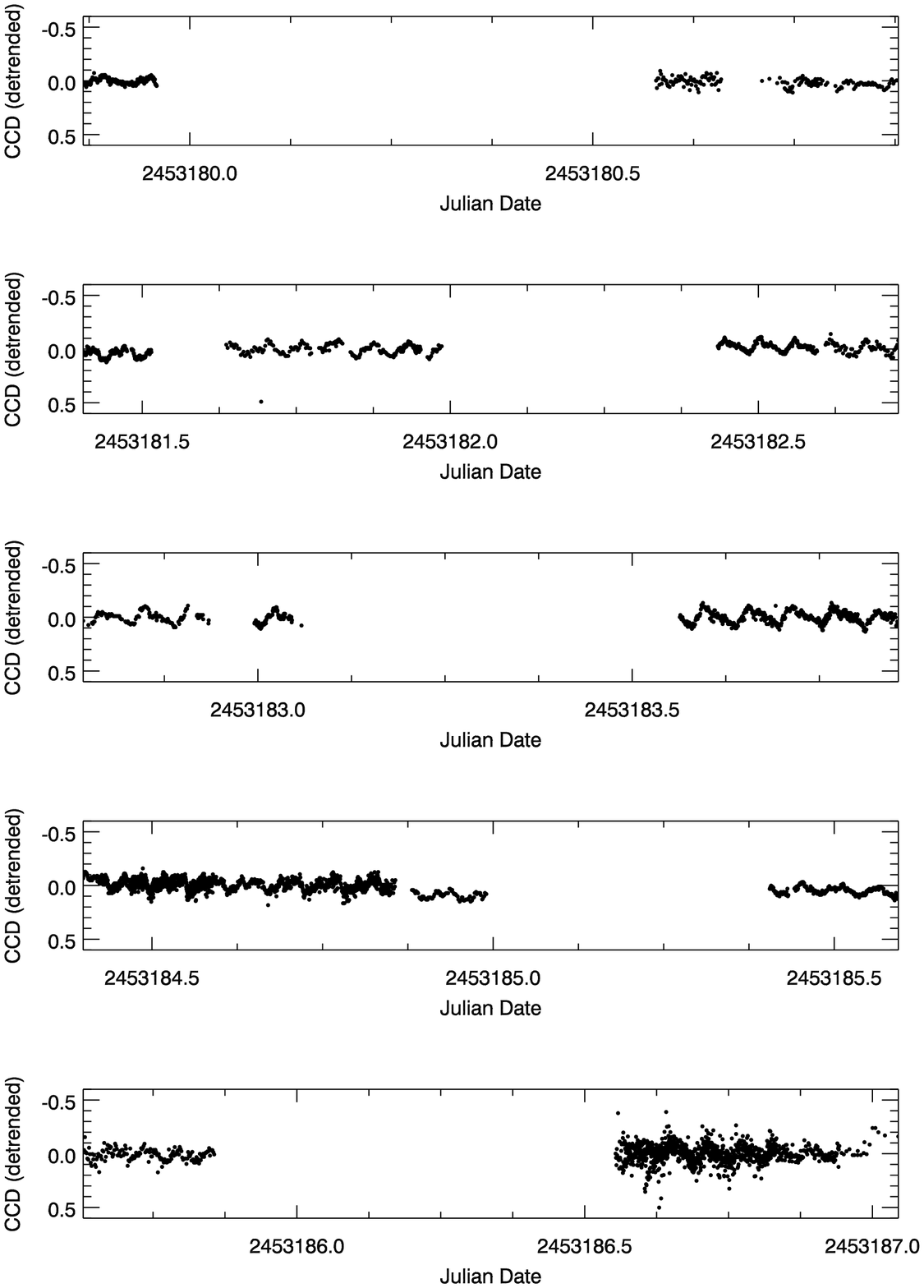]{ Superhumps during outburst. \label{six}}
\figcaption[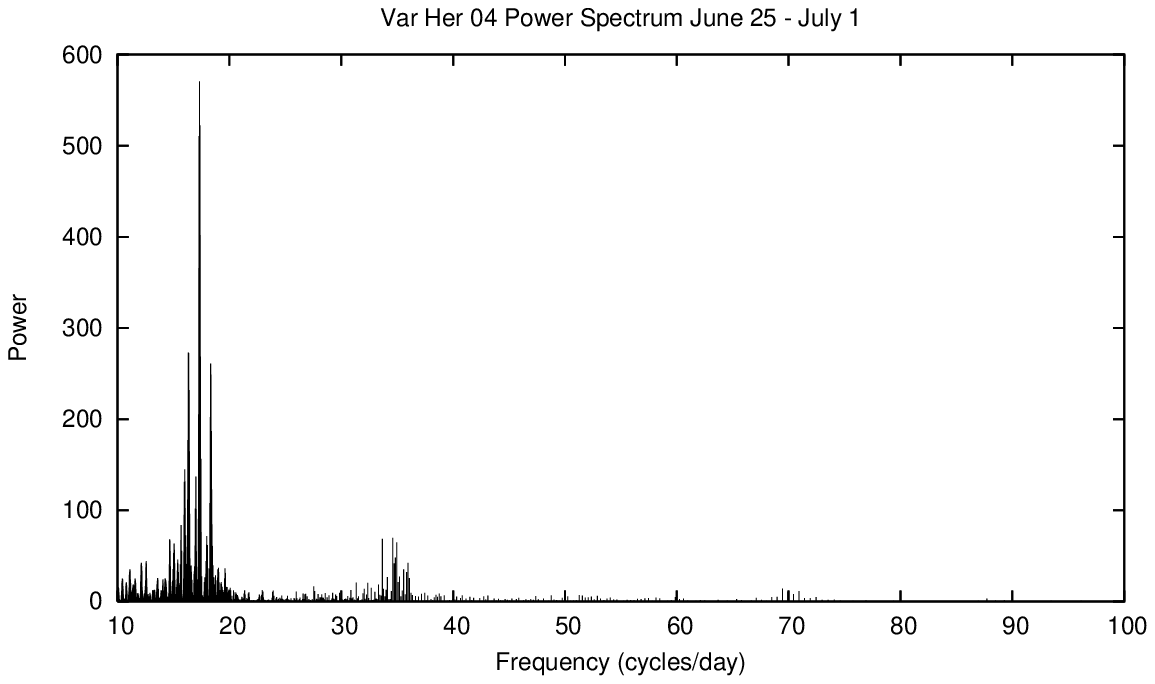]{ Power spectrum from June 23.3677 - June 30.5592 reveals a strong superhump period of 0.05778 $\pm$ 0.000001 days (17.30702 $\pm$ 0.00015 cycles/day ) .\label{seven}}
\figcaption[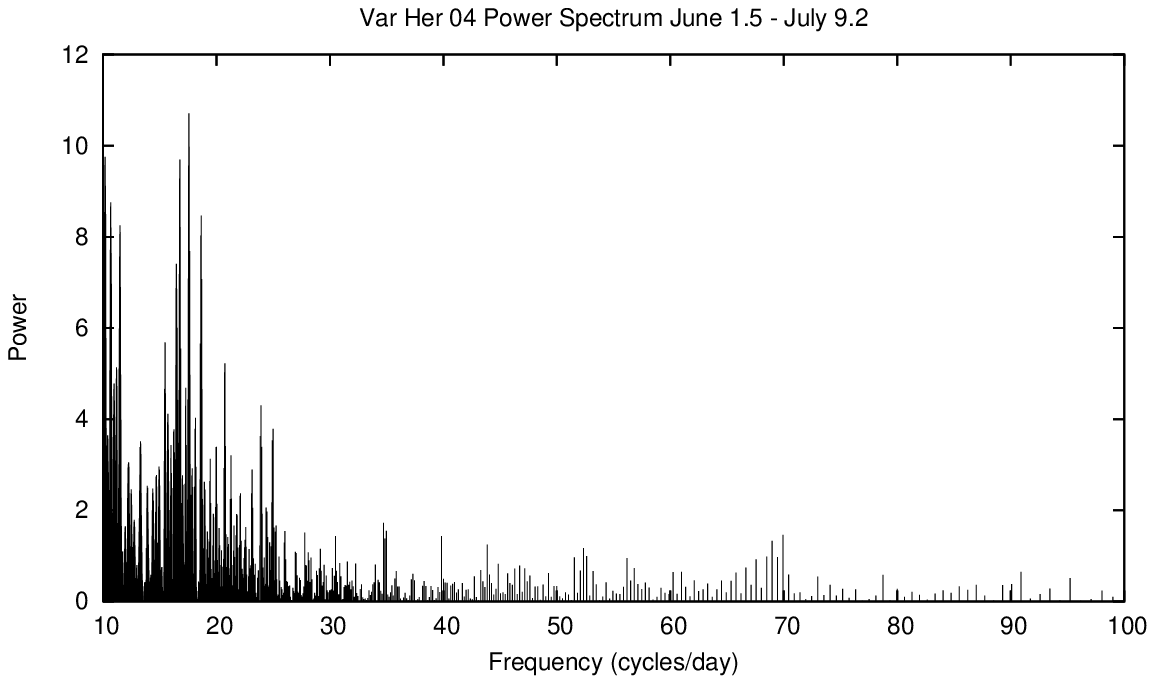]{
Quiescence power spectrum from July 1.5 - July 9.2. The orbital period of 0.056855 $\pm$ 0.000069 days (17.58860 $\pm$ 0.02139 cycles/day) emerges among noise caused by the faintess
of the object and the influence of the companion. \label{nine}}

\newpage

\begin{table}[h]\centering
\begin{tabular}{|c|c|c|c|c|} \tableline
{\it Catalog No. (USNO-A2)} & {\it Gary {\it V}}& {\it Gary {\it B-V}}& {\it Henden {\it V}}& {\it Henden {\it B-V}} \\ \tableline
 1125\_10016245  & 8.93$\pm$0.04 & 0.55$\pm$0.06 & N/A & N/A \\ \tableline
 1125\_10031765  & 10.17$\pm$0.04 & 0.54$\pm$0.06 & N/A & N/A \\ \tableline
 1125\_09987061  & 12.05$\pm$0.04 & 0.37$\pm$0.06 & N/A & N/A \\ \tableline
 1125\_10020652  & 12.77$\pm$0.04 & 0.39$\pm$0.04 & 12.781$\pm$0.011 & 0.482$\pm$0.021 \\ \tableline
 1125\_10006945  & 13.30$\pm$0.04 & 0.57$\pm$0.07 & 13.305$\pm$0.013 & 0.654$\pm$0.021 \\ \tableline
 1125\_10011471   & 14.98$\pm$0.05 & 0.64$\pm$0.10 & 15.063$\pm$0.014 & 0.746$\pm$0.016 \\ \tableline
 1125\_10011793  & 15.61$\pm$0.05 & 0.58$\pm$0.14 & 15.642$\pm$0.006 & 1.131$\pm$0.030 \\ \tableline
 1125\_10006257  & 16.52$\pm$0.15 & 0.68$\pm$0.29 & 16.595$\pm$0.017 & 1.262$\pm$0.023 \\ \tableline
\end{tabular}
\caption{Comparison star photometry used by AAVSO observers.}
\end{table}

\begin{table}[h]\centering
\begin{tabular}{|c|c|c|c|c|} \tableline
{\it Filter } & {\it Date (UT)}& {\it Magnitude}& {\it Obs.} \\ \tableline
 {\it B}  & Aug. 3, Sep. 14& 16.701$\pm$0.084 & NOFS \\ \tableline
 {\it V} & Aug. 14.3, 15.3   &  17.095$\pm$0.029  & UH\\ \tableline 
 {\it R$_c$} & Aug. 14.3, 15.3   & 17.005$\pm$0.082  & UH\\ \tableline 
 {\it I$_c$} & Aug. 14.3, 15.3   &  17.014$\pm$0.019  & UH\\ \tableline 
 {\it J} & Sep. 23.1, 26.1  &  16.885$\pm$0.064  & NOFS\\ \tableline 
 {\it H} & Sep. 23.1, 26.1   & 16.525$\pm$0.121  & NOFS\\ \tableline 
 {\it K$_p$} & Sep. 23.1, 26.1  & 16.150$\pm$0.100  & NOFS\\ \tableline 
\end{tabular}
\caption{Precision photometry of the cataclysmic variable in quiescence. NOFS observations were made 
at the U.S. Naval Observatory Flagstaff Station and UH observations made with the University of Hawaii 
2.2m on Mauna Kea.}
\end{table}

\newpage

\begin{table}[h]\centering
\begin{tabular}{|c|c|c|c|c|} \tableline
{\it Filter } & {\it Date (UT) }& {\it Magnitude}& {\it Obs.} \\ \tableline
{\it B}   & Aug. 12.2, Oct. 15  &    17.015 $\pm$0.188   & NOFS    \\ \tableline
{\it V}   & Aug. 14.3, Aug. 15, Oct. 15  &     17.098$\pm$0.061   & Combined NOFS \& UH  \\ \tableline
{\it R$_c$}  & Aug. 14.3, Aug. 15  &     16.990$\pm$0.065   & UH  \\ \tableline
{\it I$_c$}  & Aug. 14.3, Aug. 15, Oct. 15  &     16.953$\pm$0.131   & Combined NOFS \& UH  \\ \tableline
{\it J}   & Jul. 8.3   &     13.316$\pm$0.005   &  NOFS  \\ \tableline
{\it H}   & Jul. 8.3   &     12.762$\pm$0.013   &  NOFS  \\ \tableline
{\it K$_p$}  & Jul. 8.3   &     12.504$\pm$0.005   &  NOFS  \\ \tableline
{\it J}   & Aug. 4.3  &     13.410$\pm$0.005   &  NOFS \\ \tableline
{\it H}   & Aug. 4.3  &     12.830$\pm$0.006   &  NOFS  \\ \tableline
{\it K$_p$}  & Aug. 4.3  &     12.571$\pm$0.009   &  NOFS  \\ \tableline
{\it J}   & Sep. 23.1  &     13.391$\pm$0.006   &  NOFS  \\ \tableline
{\it H}   & Sep. 23.1  &     12.831$\pm$0.003   &  NOFS  \\ \tableline
{\it K$_p$}  & Sep. 23.1  &     12.560$\pm$0.011   &  NOFS  \\ \tableline
{\it J}   & Sep. 26.1  &     13.412$\pm$0.007   &  NOFS  \\ \tableline
{\it H}   & Sep. 26.1  &     12.833$\pm$$<$0.001   &  NOFS  \\ \tableline
{\it K$_p$}  & Sep. 26.1  &     12.582$\pm$0.003   &  NOFS  \\ \tableline
\end{tabular}
\caption{Photometry of the companion. Large errors are not photometric in nature but due to flickering. Note the star was brighter on July 8, coincident with the tail end of the CV's outburst. Future 
proper motion studies can determine if the companion is 
part of the same system as the variable which would make that a reflection effect. 
 }
\end{table}

\clearpage

\begin{table}[h]\centering
\begin{tabular}{|c|c|c|c|c|} \tableline 
{\it Date (UT)} & {\it Detection Magnitude (Blue)} \\ \tableline
 1932 April 21  & 10.4 \\ \tableline
 1934 October 17 & 12.2   \\ \tableline
 1939 August 16 & 14.2  \\ \tableline
 1940 August 3 & 14.4 (near plate limit)\\ \tableline
 1941 August 2 &  10.6  \\ \tableline
\end{tabular}
\caption{Previous Outbursts on Harvard Patrol Plates}
\end{table}


\begin{figure}
\epsscale{.30}
\plotone{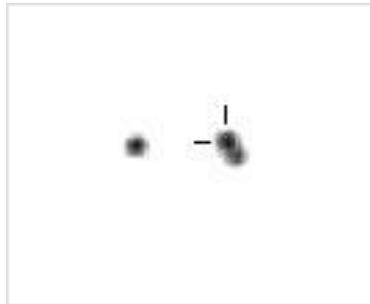}
\figurenum{1}
\caption{Var Her 04 (NW component) and companion in R$_c$ at quiescense. Image taken on Aug. 15, 2004 with the University of Hawaii 2.2m Telescope (0.2arcsec/pixel; N up, E right; image fov: 10"). 
Photometry of Ver Her 04 in quiescence is contaminated by the companion. }
\end{figure}

\begin{figure}
\epsscale{.50}
\figurenum{2}
\plotone{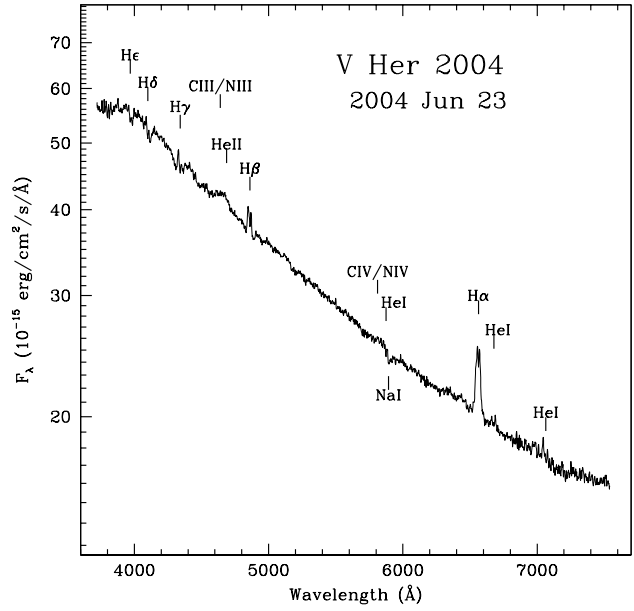}
\caption{Blue (370-750nm) spectra from Mt. Hopkins on June 23.32 reveals the typical spectrum of a cataclysmic variable.}
\end{figure}


\begin{figure}
\epsscale{.50}
\figurenum{3}
\plotone{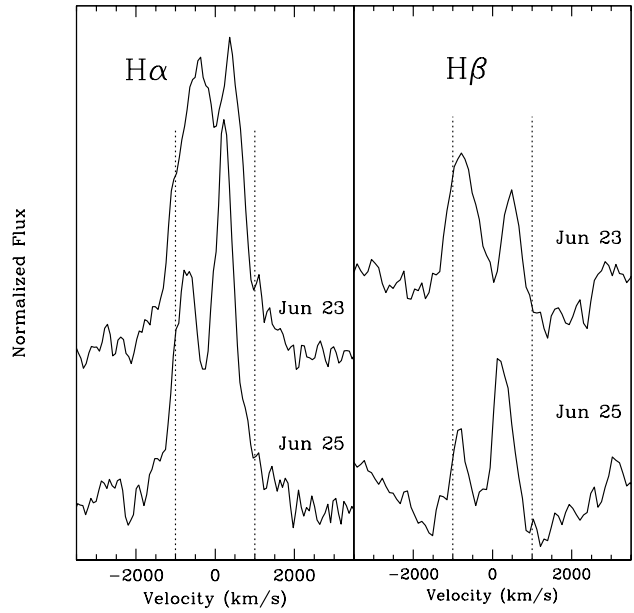}
\caption{Comparison of Ha and Hb features in the Mt. Hopkins spectra taken on June 23.32 and June 25.00.}
\end{figure}
 
\begin{figure}
\figurenum{4}
\epsscale{.90}
\plotone{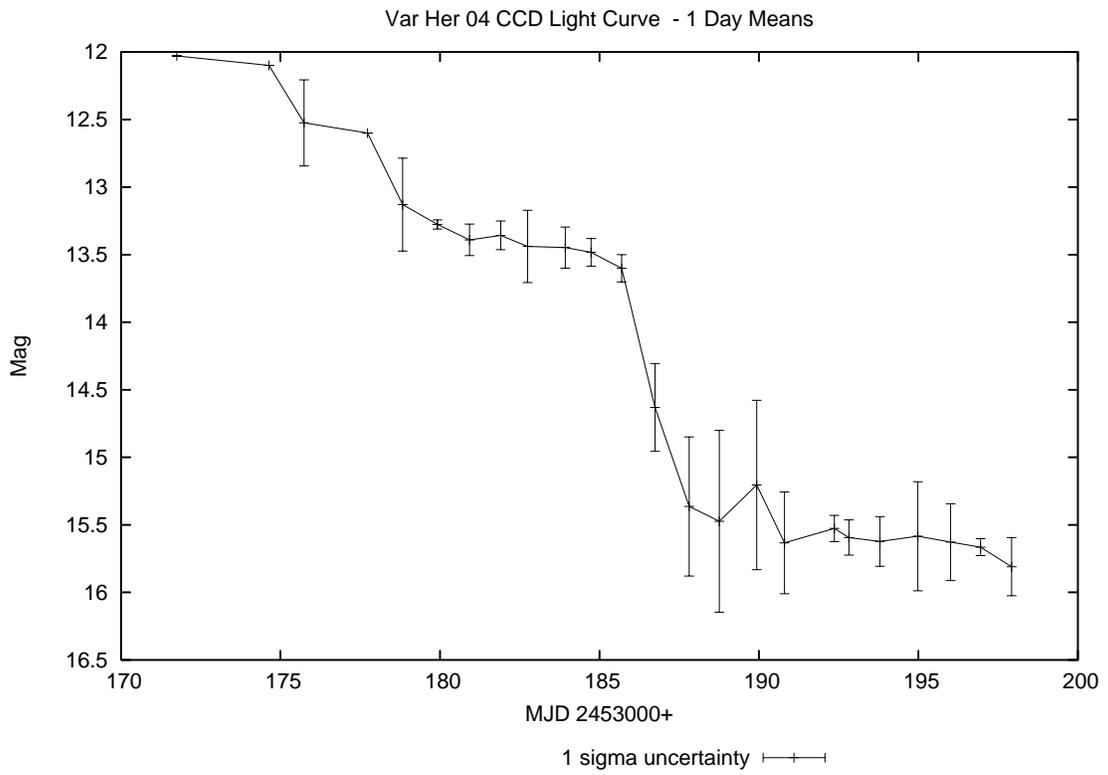}
\caption{Var Her 04's Light Curve. Data averaged from AAVSO sources. Combining faint unfiltered observations leads to the larger errors
      at late time. }
\end{figure}

\begin{figure}
\figurenum{5}
\epsscale{.90}
\plotone{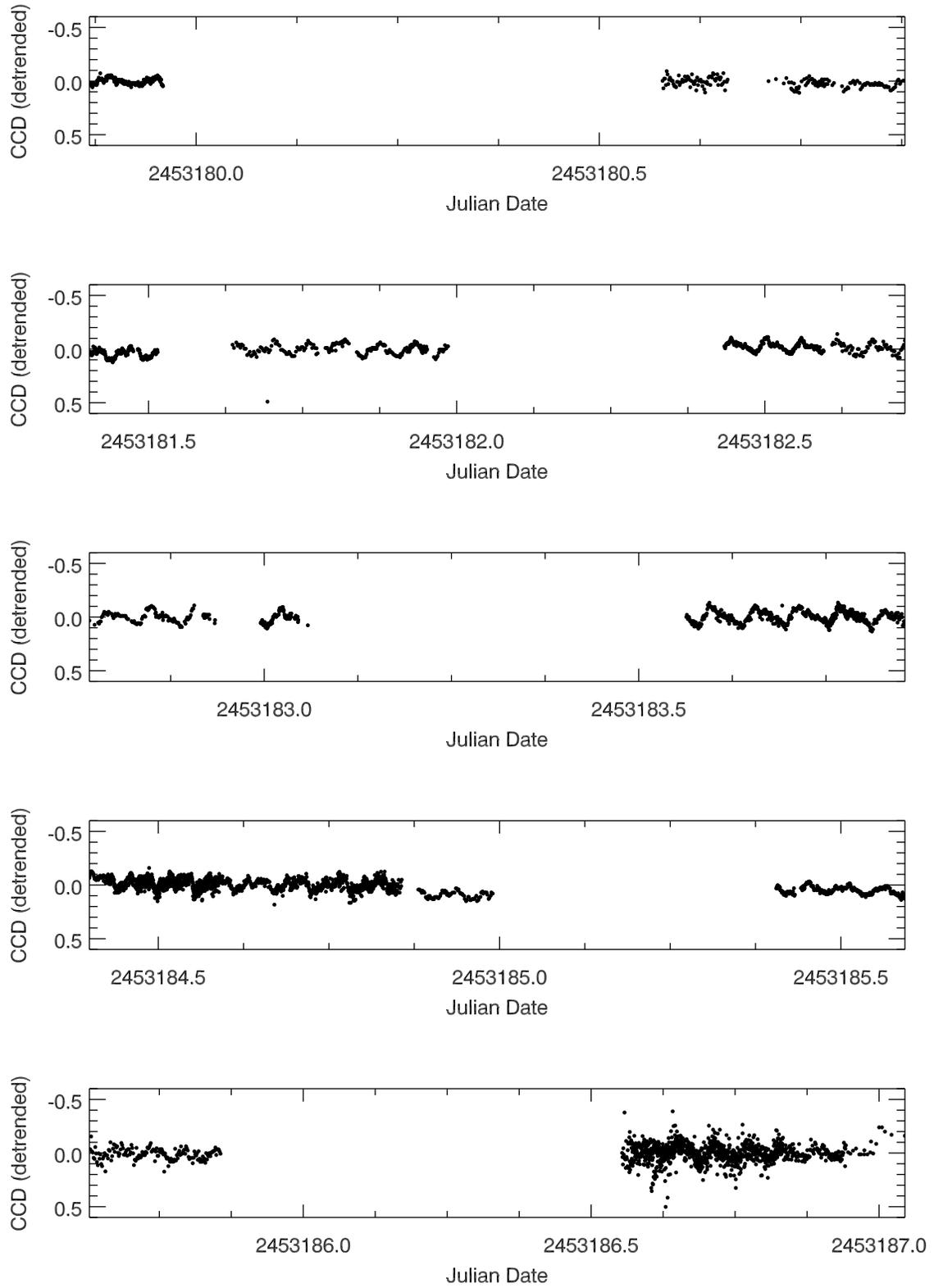}
\caption{Superhumps during the outburst.}
\end{figure}

\begin{figure}
\figurenum{6}
\epsscale{.50}
\plotone{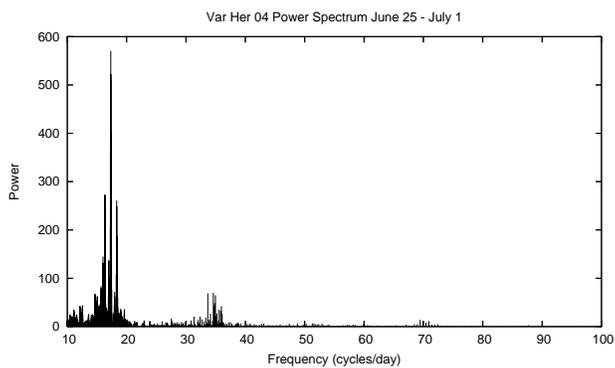}
\caption{Power spectrum from June 23.3677 - June 30.5592 reveals a strong superhump period of 0.05778 $\pm$ 0.000001 days (17.30702 $\pm$ 0.00015 cycles/day ).}
\end{figure}

\begin{figure}
\figurenum{7}
\epsscale{.50}
\plotone{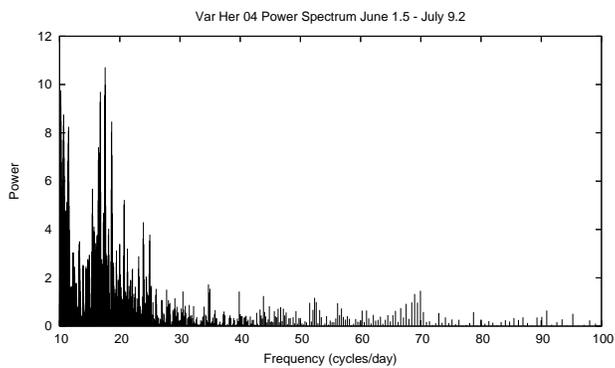}
\caption{Quiescence power spectrum from July 1.5 - July 9.2. The orbital period of 0.056855 $\pm$ 0.000069 days (17.58860 $\pm$ 0.02139 cycles/day) emerges among noise caused by the faintess 
of the object and the influence of the companion. }
\end{figure}


\newpage 

\begin{thebibliography}{}

\bibitem[Brosch et al.(1980)]{bro80} Brosch, N. et al. 1980, \apjl, 236, L29
\bibitem[Fabricant et al.(1998)]{fab98} Fabricant, D., Cheimets, P., Caldwell, N., \& Geary, J. 1998, \pasp, 110, 79
\bibitem[Ferraz-Mello(1981)]{fer81} Ferraz-Mello, S. 1981, \aj, 86, 619
\bibitem[Filippenko(1982)]{fil82} Filippenko, A. V. 1982, \pasp, 94, 715
\bibitem[Foster(1995)]{fos95} Foster, G. 1995, \aj, 109, 1889. 
\bibitem[Green et al.(2004)]{gre04} Green, D. et al. 2004, \iaucirc, 8374
\bibitem[Landolt(1992)]{lan92} Landolt, A. 1992, \aj, 104, 340
\bibitem[Matheson et al.(2000)]{mat00} Matheson, T., Filippenko,  A. V., Ho, L. C., Barth, A. J., \& Leonard, D. C. 2000, \aj, 120, 1499
\bibitem[Oke \& Gunn (1983)]{oke83} Oke, J. B., \& Gunn J. E. 1983, \apj, 266, 713
\bibitem[Patterson(1998)]{pat98} Patterson, J. 1998, \pasp, 110, 1132 
\bibitem[Pojmanski(2002)]{poj02} Pojmanski, G. 2002, \actaa, 52, 396 
\bibitem[Stone(1977)]{sto77} Stone, R. P. S. 1977, \apj, 218, 767
\bibitem[Voges et al.(1999)]{vog99} Voges, W. et al. 1999, \aap, 349, 389
\bibitem[Wade \& Horne(1988)]{wad88} Wade, R. A., \& Horne,
    K. D. 1988, \apj, 324, 411
\bibitem[Wozniak et al.(2004)]{woz04} Wozniak, P. R., et al. 2004. \aj,  127, 2436
\end{thebibliography}
\end{document}